\documentclass[aps,prl,preprint,superscriptaddress]{revtex4}
\usepackage{graphicx}

\def\eq{\begin{equation}}
\def\en{\end{equation}}
\def\eqa{\begin{eqnarray}}
\def\ena{\end{eqnarray}}

\newcommand{\dif}{\mathrm{d}}

\def\Tr{\mathrm{Tr}}
\def\expval#1{\langle #1 \rangle}
\def\expvalc#1{\langle #1 \rangle_{c}}
\def\bra#1{\langle#1|}
\def\ket#1{|#1\rangle}
\def\Tbulk{T^{\mathit bulk}}
\def\Tsbulk{T_{\mathit bulk}}
\def\Thetabulk{\Theta_{\mathit bulk}}
\def\textint{\int}
\def\str{\mathit{s}}
\def\betastr{\beta_{\str}}
\def\gstr{g_{\str}}
\def\Gstr{G^{\str}}
\def\Zstr{Z_{\str}}
\def\idop{\mathbf{1}}

\begin{document}

\preprint{RU-NHETC-2003-39}
\preprint{hep-th/0312197}

\title{On the Boundary Entropy of One-dimensional Quantum Systems at
Low Temperature}

\author{Daniel Friedan}
\email{friedan@physics.rutgers.edu}
\affiliation{Department of Physics and Astronomy,
Rutgers, The State University of New Jersey,
Piscataway, New Jersey 08854-8019 U.S.A.}
\affiliation{The Science Institute, The University of Iceland,
Reykjavik, Iceland}

\author{Anatoly Konechny}
\email{tolya@phys.huji.ac.il}
\affiliation{Racah Institute of Physics, The Hebrew University,
Jerusalem 91904, Israel}

\date{\today}

\begin{abstract}
The boundary $\beta$-function generates the renormalization group
acting on the universality classes of one-dimensional quantum
systems with boundary which are critical in the bulk but not
critical at the boundary.  We prove a gradient formula for the
boundary $\beta$-function, expressing it as the gradient of the
boundary entropy $s$ at fixed non-zero temperature.  The gradient
formula implies that $s$ decreases under renormalization except at
critical points (where it stays constant).  At a critical point, the
number $\exp(s)$ is the ``ground-state degeneracy,'' $g$, of Affleck
and Ludwig, so we have proved their long-standing conjecture that
$g$ decreases under renormalization, from critical point to critical
point.  The gradient formula also implies that $s$ decreases with
temperature except at critical points, where it is independent of
temperature.  The boundary thermodynamic energy $u$ then also
decreases with temperature.  It remains open whether the boundary
entropy of a 1-d quantum system is always bounded below.  If $s$ is
bounded below, then $u$ is also bounded below.
\end{abstract}


\maketitle

\section{Introduction}
The logarithm of the partition function $\ln Z = \ln \Tr \exp(-\beta
H)$ for a one-dimensional quantum critical system with a boundary
takes the universal form\cite{AL1} $ (c\pi/6) (L/\beta)+ \ln g$
where $H$ is the hamiltonian, $\beta = 1/T$ the inverse temperature,
$L\gg \beta$ is the length, $c$ is the numerical coefficient of the
bulk conformal anomaly, and $g$ is the< ``universal noninteger
ground-state degeneracy'' at the boundary (using natural units in
which $\hbar = k=v=1$, $v$ being the velocity of ``light'').  This
formula applies in the limit of large $L$.  The number $g$ is an
invariant of the universality class of the critical boundary
condition.  It was conjectured that $g$ decreases from critical
point to critical point under renormalization\cite{AL1,AL2}.

For a 1-d quantum system that is critical in the bulk but is
\emph{not} critical at the boundary, the logarithm of the partition
function at low temperature can be written in the form $\ln Z =
(c\pi/6) (L/\beta) + \ln z_{L}$ and the boundary partion function
$z$ can be defined as $\lim_{L\rightarrow\infty} z_{L}$.  That is, the
partition function takes the universal form
\eq \label{eq:partitionfn}
\ln Z = (c\pi/6) (L/\beta) + \ln z
\en
up to corrections which vanish in the limit $L\rightarrow\infty$.

Quantum critical points occur at zero temperature.  At temperature
$T>0$, the correlation functions are of course not scale-invariant,
but decay exponentially at distance scale $\beta=1/T$.
Nevertheless, it is meaningful to talk of a bulk critical system at
temperature $T>0$.  The low energy degrees of freedom and their
coupling constants can be identified in the $T=0$ critical system,
which is described by a 1+1
dimensional quantum field theory.  That quantum field theory, at
temperature $T>0$, describes the quantum critical system at
temperature $T$.  It is
possible to identify temperature-independent coupling constants within
the critical system.  These coupling constants can be held to
their critical values when the temperature is above zero.
Quantum critical phenomena are special in this respect.
In generic critical phenomena,
which have non-zero
critical temperatures, the temperature is just one of the coupling
constants.

Given that the bulk system is critical and that $L=\infty$, the only
dimensionful parameter is the temperature $T$.  The logarithm of the
boundary partition function is thus a function $\ln z(\mu\beta)$ that
depends only on
the temperature, in units of $\mu$, where $\mu$ is a small
temperature that sets the renormalization scale (or, equivalently, a
small energy or inverse time or inverse distance).  The total
entropy then takes the universal form
\eq
S=(1- \beta\partial/\partial \beta)\ln Z
= (c\pi/3) (L/\beta)+ s(\mu \beta)
\en
where $s(\mu\beta)= (1- \beta\partial/\partial \beta)\ln z$ is the boundary
entropy.  At a critical point, $s$ is equal to the constant $\ln g$.

We prove here a gradient formula
\eq \label{eq:gradient}
g_{ab}(\lambda) \beta^{b}(\lambda) = - \partial s/\partial\lambda^{a}
\en
where the $\lambda^{a}$ form a complete set of boundary coupling
constants, $g_{ab}(\lambda)$ is a certain metric on the space of all
boundary conditions, and $\beta^{a}(\lambda)$ is the boundary
$\beta$-function.
It follows directly from the gradient formula that
$
\mu \partial s/\partial \mu = \beta^{a}\partial_a s =
- g_{ab}\beta^{a}\beta^{b}
$
so $s$ decreases under the renormalization group except where
$\beta^{a}=0$, at the critical points.  The gradient formula
eliminates the possibility of esoteric asymptotic behavior under
renormalization.  Recurring trajectories such as limit cycles are
excluded.  The $g$-conjecture for the rg flows between critical
points is a corollary of the gradient formula.

The gradient formula implies equally that the boundary entropy
decreases with temperature, $\beta \partial s/\partial\beta =
\mu \partial s/\partial\mu < 0$.  The total entropy $S$ obviously
decreases with temperature, because $\partial S/\partial\beta = -
\beta \expval{(H-\expval{H})^{2} }$.  However, the decrease of the
bulk contribution $(\pi/3) (L/\beta)c$ masks the change in $s$, so
it is not obvious that the boundary entropy by itself must decrease
with temperature.  The gradient formula implies that it does.
It follows that the thermodynamic boundary energy also decreases
with temperature, $ \mu \beta \partial u/\partial \beta = \partial
s/\partial \beta <0$.

Complete control over the possible behavior at asymptotically low
temperature is still lacking, because we do not prove that $s$ is
bounded below.  If $s$ is bounded below, then the system must go to
a critical point at zero temperature.  Of course, the total entropy
$S$ of any system is bounded below, as long as the system is of
finite size.  So, for any finite size $L$, $s_{L}= S- (c\pi/3)
(L/\beta)$ is bounded below, as $T\rightarrow 0$.  However, the
lower bound can descend without limit as $L\rightarrow\infty$, so $s
= \lim_{L\rightarrow\infty}s_{L}$ is not necessarily bounded below,
as $T\rightarrow 0$.  It still remains to be proved that
$s(\mu\beta)$ is bounded below, as $T\rightarrow 0$.  If the
boundary entropy is bounded below, then the boundary energy is also
bounded below.

The gradient formula that we prove is mathematically equivalent to a
gradient formula conjectured in string
theory\cite{Witt1,Witt2,Shat1,Shat2,KMM}.  Evidence was given for
the string theory conjecture\cite{Witt2,Shat2,AK}, but the formula
was never proved.  It has been claimed that a proof was given in
Ref.~\cite{KMM}, but it was assumed there that the boundary
$\beta$-function $\beta^{a}(\lambda)$ is linear in the coupling
constants $\lambda^{a}$.  This is an invalid assumption.  The
$\beta$-function cannot be linearized when there are marginally relevant
couplings or, more generally, whenever resonance
conditions occur (as discussed, for example, in Ref.~\cite{Shat2}).
Moreover, the conjectured string gradient formula is expressed in
un-physical quantities, in terms of un-normalized correlation
functions.  Our contribution is to express the gradient formula in
terms of normalized correlation functions and the boundary entropy,
which are physical quantities, and to prove the formula using
physical properties of the 1-d quantum system.  Some of the ideas
used in the proof can be found in the string theory
work\cite{Witt1,Shat2}.  The re-writing of the conjectured string
gradient formula is based on an idea that is implicit in
Ref.~\cite{KMM} and was mentioned explicitly to us\cite{Moor1}.
Here, to avoid distracting from the physical meaning, we first prove
the gradient formula in physical terms, and only afterwards explain
the connection to the string conjecture.

The proof of the gradient formula applies to \emph{all} local 1d quantum systems.
It uses only the basic principles of quantum mechanics and locality.
The gradient formula must therefore hold in every
local 1d quantum mechanical model.
The point of proving a result such as the gradient formula is to
give reliable theoretical information about what is physically
possible.  For instance, when building devices out of low
temperature 1-d quantum systems joined at boundaries, it will be
useful to know in advance, with certainty, what kinds of boundary
behaviors are possible.  It will be useful to know that the boundary
must always behave as a thermodynamic system, except that it does
not obey the third law.  Proof also reveals what must be done to
evade the theoretical limits.  The gradient formula itself is not
likely to be avoidable, since the proof depends only on the basic
principles of quantum mechanics and renormalization, assuming only
the existence of a local stress-energy tensor, which is assured by
microscopic locality.  Rather, attention is directed towards exotic
systems, where the metric $g_{ab}(\lambda)$ degenerates, or where
$s$ is infinite\cite{DFnotes,Tseng}, or even where $s$ might not be
bounded below, if this cannot be proved impossible.  A lower bound
on $s$ would have to depend on the details of the bulk system.  The
bound could not be uniform, not a function of $c$ alone.  This can
be seen in the critical $c=1$ gaussian model, where the values of
$g$ depend on the marginal coupling constant of the bulk model, and
can become arbitrarily close to zero\cite{ERS}.

\section{The stress-energy tensor in the presence of a boundary}

\begin{figure}[!h] 
\centering
\includegraphics[width=200pt]{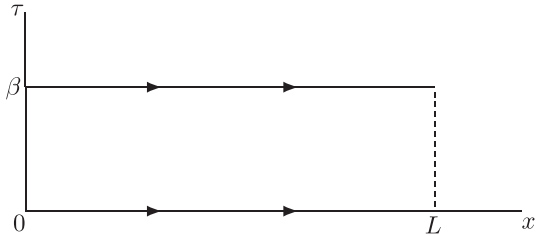}
\caption{The space-time
is periodic in imaginary time $\tau$, with period $\beta$.  The
boundary is at $x=0$.}
\label{fig:1}
\end{figure}

The equilibrium observables of the system live on the cylindrical
euclidean spacetime, periodic in euclidean time with period $\beta$
(see Fig.~\ref{fig:1}).
The spacetime coordinates are $x^{\mu}=(x,\tau)$, $0\le x < L$,
$\tau\sim \tau+\beta$.  The boundary is at $x=0$.

The stress-energy tensor $T_{\mu \nu}(x,\tau)$ expresses the
response of the system to an infinitesimal local variation of the
metric, $g_{\mu \nu} \rightarrow \delta_{\mu \nu} + \delta g_{\mu
\nu}(x,\tau)$,
\eq
\delta \ln Z = (1/2)
\textint\!\!\!\textint \mu^{2}\dif\tau \dif x \,
\expval{\delta g_{\mu \nu} T^{\mu \nu}(x,\tau)}
\,.
\en
We specialize to 1+1 dimensions
the general analysis of the stress-energy tensor in space-times
with boundary\cite{McAO}.
The stress-energy tensor can be written as a bulk part plus a
boundary part
\[
\delta \ln Z =
(1/2)
\textint\!\!\!\textint \mu^{2}\dif\tau \dif x \,
\expval{\delta g_{\mu \nu}\Tsbulk^{\mu\nu}(x,\tau)}
+(1/2) \textint_{0}^{\beta} \mu \dif\tau \,
\expval{\delta g_{\mu \nu}(0,\tau) t^{\mu\nu}(\tau)}
\]
\eq
T_{\mu\nu} = \Tbulk_{\mu\nu}(x,\tau)
+ \delta(\mu x) t_{\mu\nu}(\tau)
\, .
\en
There could also be a boundary operator proportional to
$\delta'(x)\idop$, but the identity operator
makes no contribution to connected correlation functions,
so we can ignore it.

The conservation equations follow from invariance of the physics
under localized coordinate reparametrizations
$\delta x^{\mu} = v^{\mu} (x,\tau)$ where
the vector field $v^{\mu}$ is tangent
to the boundary, i.e.
$v^{x}(0,\tau)=0$.
The coordinate reparametrization
is equivalent to a change in the metric tensor
$
\delta g_{\mu \nu} = \partial_{\mu}v_{\nu} + \partial_{\nu}v_{\mu}
\, .
$
Plugging this into the formula for $\delta \ln Z$
and setting the variation to zero
we obtain, after integration by parts, the bulk conservation equation
$\partial^{\mu}\Tbulk_{\mu\nu}=0$
and also
\[
\textint
\mu \dif \tau \, \left (
\mu   \Tbulk_{x \tau} v^{\tau} -
t_{\mu \nu} \partial^{\mu}v^{\nu}
\right )= 0\, .
\]
at the boundary, which is equivalent to the boundary conservation equations
$t_{xx}=t_{x\tau}=t_{\tau x}=0$ and
\eq \label{eq:b_con2}
\mu \Tbulk_{x\tau}(0,\tau) +  \partial_{\tau} \theta(\tau) = 0
\en
where $\theta(\tau) \equiv t_{\tau\tau}(\tau)$.
The boundary operator $\theta$ was described
in Ref.~\cite{GhZ}.

The trace of the stress-tensor is
\begin{equation}
T_{\mu}^{\mu}= \Theta(x,\tau) =
\Thetabulk(x,\tau)
+
\delta(\mu x)\theta(\tau) \, .
\end{equation}
The system is critical in the bulk, so $\Thetabulk(x,\tau)=0$
up to contact terms.
The full trace is $\Theta=\delta(\mu x)\theta(\tau)$,
entirely a boundary operator.

The space of boundary conditions is parameterized by the coupling
constants $\lambda^{a}$
which couple to the renormalized local
boundary fields $\phi_{a}$
\[
\partial_a \ln Z
=
\textint \mu\dif\tau \, \expval{\phi_{a}(\tau)}
\, .
\]
The boundary trace $\theta$ can be decomposed
into a linear combination
of the boundary fields and the identity operator
\eq
\theta\ = \beta^{a}(\lambda) \phi_{a}
+ h(\lambda) \idop
\en
where the coefficients $\beta^{a}(\lambda)$ are the boundary
$\beta$-functions.  We will not have to worry about the term
$h(\lambda)\idop$, because $\theta$ will only
appear within connected correlation functions.

The foregoing are operator statements.
In correlation functions, the stress-energy tensor
will also have contact terms.
The generator of dilatations
$\delta g_{\mu\nu}= 2(\delta \mu/\mu)\delta_{\mu\nu}$
is
$
\textint\!\!\!\textint \mu^{2}\dif \tau \dif x \,
\Theta (x,\tau)
$
so the renormalization group equation for
$\ln Z$ is
\eqa \label{eq:rgz}
(\mu \partial/\partial \mu) \ln Z &=&
\textint\!\!\!\textint \mu^{2}\dif \tau \dif x \,
 \expval{\Theta(x,\tau) } \nonumber\\
&=& \textint\!\!\!\textint \mu^{2}\dif \tau \dif x \,
 \expvalc{\delta(x)\theta(\tau)+ \Thetabulk(x,\tau)} \nonumber\\
&=& \beta^{a} \partial_{a} \ln Z + \mu \beta h(\lambda)
\, .
\ena
For the one-point functions,
\eqa
(\mu \partial/\partial \mu) \expval{\phi_{b}(\tau_{1})}
&=& \textint\!\!\!\textint \mu^{2}\dif \tau \dif x \,
 \expvalc{\Theta(x,\tau) \phi_{b}(\tau_{1})} \nonumber\\
&=& \textint\!\!\!\textint \mu^{2}\dif \tau \dif x \,
 \expvalc{\left [\delta(x)\theta(\tau)+ \Thetabulk(x,\tau)
\right ] \phi_{b}(\tau_{1})}
 \nonumber\\
&=& \beta^{a} \partial_{a} \expval{\phi_{b}(\tau_{1})}
+(\gamma^{a}_{b}-\delta^{a}_{b})\expval{\phi_{a}(\tau_{1})}
\ena
where the coefficients $\gamma^{a}_{b}-\delta^{a}_{b}$ come from
contact terms of $\Thetabulk$ and $\theta$ with $\phi_{a}$.  Because
of the contact terms, $\Thetabulk$ cannot be omitted.
The identity
$\gamma^{a}_{b}=\partial_{b}\beta^{a}$ follows from
$[\mu\partial/\partial\mu, \partial_{a}]=0$, which in turn follows
from the definition of the $\lambda^{a}$ as the coupling constants
renormalized at scale $\mu$.

We will need one last property of the stress-energy tensor, that
$\Tbulk_{\mu\nu}(x,\tau)$ decays as $\exp(-4\pi x/\beta)$ in
connected correlation functions far from the boundary.  When $x$ is
far from the boundary, $\Tbulk_{\mu\nu}(x,\tau)$ behaves as in the
bulk theory without boundary.  The exponential decay condition
expresses the conformal invariance of the bulk critical system.
It is derived using the interpretation of Fig.~\ref{fig:1}
in which space and imaginary time are exchanged.  Space becomes a
circle of length $\beta$.  The correlation functions become the
expectation values $\bra{B} \cdots \ket{0}$ where $\ket{0}$ is the
ground state of the bulk critical system on the circle, and
$\bra{B}$ is the state representing the boundary condition.  Using
the complex coordinate $w= 2\pi (x+i\tau)/\beta$, the bulk
stress-energy tensor takes the form
$\Tbulk_{xx}=\Tbulk_{\tau\tau}=T(w)+\bar T(\bar w)$,
$\Tbulk_{x\tau}= i(T(w)-\bar T(\bar w))$, where $T(w) = \sum_{n}
L_{n}\exp(-nw)$ and $\bar T(\bar w) = \sum_{n}\bar L_{n}\exp(-n\bar
w) $, $n$ ranging from ${-}\infty$ to ${+}\infty$, the $L_{n}$ and
$\bar L_{n}$ being the Virasoro operators.  Bulk conformal
invariance means $L_{1}\ket{0}=\bar L_{1}\ket{0}=0$.  Therefore,
$T(w)\sim \exp(-2w)$ and $T(\bar w)\sim \exp(-2\bar w)$ in connected
correlation functions, far from the boundary.  So
$\Tbulk_{\mu\nu}(x,\tau)\sim\exp(-4\pi x/\beta)$.

\section{The proof}
We prove the gradient formula, Eq.~\ref{eq:gradient}, with the
metric on the space of boundary conditions given by
\eq \label{metric}
g_{ab}(\lambda) = \textint\mu \dif \tau_{1}
\textint\mu \dif \tau
\,  \langle \phi_{a}(\tau_{1})\phi_{b}(\tau )\rangle_{c}
\left (1-\cos\left[2\pi(\tau -\tau_{1})/\beta\right] \right) \, .
\en
This is essentially the metric proposed in Ref.~\cite{Witt1}, except
that Ref.~\cite{Witt1} used the un-normalized, full two-point
function, while we use the normalized, connected two-point function.
Because we are using the connected two-point function,
we can write
\begin{equation} \label{LHS}
g_{ab}\beta^{b} = \textint\mu \dif \tau_{1}
\textint\mu \dif \tau \,
\langle \phi_{a}(\tau_{1})\theta(\tau )\rangle_{c}
\left (1-\cos\left[2\pi(\tau -\tau_{1})/\beta\right] \right)
\, .
\end{equation}
The identity component of $\theta$ makes no contribution
to the connected two-point function.
Let us deal with the term containing the cosine:
\eqa \label{A}
A_{a}(\tau_{1})&\equiv&
\textint\mu \dif \tau
\langle \phi_{a}(\tau_{1})\theta(\tau)\rangle_{c}
\left (-\cos\left[2\pi(\tau -\tau_{1})/\beta\right] \right)
\nonumber\\
&=&
\textint \mu\dif\tau
\langle \phi_{a}(\tau_{1}) \partial_{\tau}\theta(\tau)\rangle_{c}
(\beta/2\pi) \sin\left[2\pi(\tau -\tau_{1})/\beta\right]
\ena
We have integrated by parts on the boundary
to obtain the second equation.
The correlation functions are distributions,
so integration by parts is justified.
By the boundary conservation law Eq.~\ref{eq:b_con2},
\eq
A_{a}(\tau_{1}) =
\textint \mu \dif \tau \, \langle \phi_{a}(\tau_{1})
\mu\Tbulk_{x\tau}(\tau)\rangle_{c} (-2) v^{\tau}(0,\tau)
\en
where we define $v^{\tau}(0,\tau) \equiv
(\beta/4\pi)\sin\left[2\pi(\tau -\tau_{1})/\beta\right]$ as a
tangent vector field on the boundary.
Next, we extend the boundary vector field $v^{\tau}(0,\tau)$
to a conformal
Killing vector field $v^{\mu}( x, \tau)$ in the bulk.
That is, $v^{x}(0, \tau)=0$ and
$
\partial_{\mu}v_{\nu} + \partial_{\nu}v_{\mu} =
g_{\mu \nu} \partial_{\sigma}v^{\sigma}
$.
Such a vector field is most easily found as an analytic vector field
$v^{w}=(2\pi/\beta)
(v^{x}+iv^{\tau})$
in the
complex coordinate $w= 2\pi (x+i\tau)/\beta$,
\eqa
v^{w}&=&(2\pi/\beta)
(v^{x}+iv^{\tau})= \left [
\exp(w - w_{1}) - \exp(-w + w_{1})\right ] /4  \nonumber\\
\bar v^{\bar w}
&=& (2\pi/\beta)(v^{x}-iv^{\tau}) =
\left [
\exp(\bar w - \bar w_{1}) -\exp(-\bar w + \bar w_{1})
\right ] /4
\, .
\nonumber
\ena
Then
\[
\partial_{\sigma}v^{\sigma} = \partial_{w}v^{w} +
\partial_{\bar w}\bar v^{\bar w}
= \cos\left[2\pi(\tau - \tau_{1})/\beta\right]
\cosh(2\pi x/\beta)
\,.
\]
Now we integrate by parts in the bulk,
using the bulk conservation equation,
to obtain
\eq
A_{a}(\tau_{1})=
\textint\!\!\!\textint
\mu^{2} \dif \tau \dif x \, \langle \phi_{a}(\tau_{1})
\Tbulk_{\mu\nu}(x,\tau)\rangle_{c}
(\partial^{\mu}v^{\nu}+\partial^{\nu}v^{\mu})
\,.
\en
There is no boundary term at large $x$ because of the decay
condition $\Tbulk_{\mu\nu}(x,\tau)\sim\exp(-4\pi x/\beta)$.
Then we use
the fact that $v^{\mu}$ is a conformal Killing vector
to write
\eq
A_{a}(\tau_{1})=
\textint\!\!\!\textint
\mu^{2} \dif \tau \dif x \,
\langle \phi_{a}(\tau_{1}) \Thetabulk(x,\tau)\rangle_{c}
\partial_{\sigma}v^{\sigma}
\,.
\en
Finally, we can approximate $\partial_{\sigma}v^{\sigma}
\sim 1$, because $\Thetabulk=0$ except for contact terms.
The error term is
\[
\textint\!\!\textint \mu^{2} \dif \tau \dif x \,
\langle \phi_{a}(\tau_{1}) \Thetabulk(x, \tau)\rangle_{c}
(\partial_{\sigma}v^{\sigma} - 1)
\, .
\]
The boundary operator $\phi_{a}(\tau_{1})$ is renormalizable, and
$\Thetabulk$ has dimension $2$, so the most singular contact terms
in the two-point function are of the form $\delta(x)\delta'(\tau -
\tau_{1})$ and $\delta'(x)\delta(\tau - \tau_{1})$.  But
$\partial_{\sigma}v^{\sigma}(x,\tau) - 1$ vanishes to second order
at $x=0$, $\tau=\tau_{1}$, so there is no error.
Thus
\eq
\label{eq:Aa}
A_{a}(\tau_{1}) =
\textint\!\!\!\textint
\mu^{2} \dif \tau \dif x \,
\langle \phi_{a}(\tau_{1}) \Thetabulk(x,\tau)\rangle_{c} \, .
\en

Using Eq.~\ref{eq:Aa} in Eq.~\ref{LHS},
we arrive at
\eqa \label{LHS2}
g_{ab}\beta^{b} &=&
\textint\mu \dif \tau_{1}
\textint\!\!\textint \mu^{2}\dif\tau \dif x \,
 \langle \phi_{a}(\tau_{1})
 \left [ \delta(\mu x)\theta( \tau) + \Thetabulk(x,\tau) \right ]
\rangle_{c} \nonumber\\
&=&
\textint\mu \dif \tau_{1}
\textint\!\!\textint \mu^{2}\dif\tau \dif x \,
 \langle \phi_{a}(\tau_{1})
 \Theta(x,\tau) \rangle_{c} \nonumber\\
&=&
\textint\mu \dif \tau_{1}
 (\mu \partial/\partial \mu )
\expval{\phi_{a}(\tau_{1})} \nonumber\\
&=&
(\mu \partial/\partial \mu - 1)
\partial_{a} \ln Z \nonumber\\
&=&
\partial_{a}
(\mu \partial/\partial \mu - 1)
\ln Z \nonumber\\
&=&
- \partial_{a} s
\ena
which is the gradient formula.

\section{Comments on the gradient formula}
Each element of the gradient formula is covariant under renormalization.  The
boundary entropy $s$ is covariant under renormalization,
$\mu\partial s/\partial\mu = \beta^{a}\partial_{a}s$, even though
the partition function is not (see Eq.~\ref{eq:rgz}).
Using Eq.~\ref{eq:rgz},
\eqa
(\mu\partial
/\partial\mu
-\beta^{a}\partial_{a})
s
&=& (\mu\partial /\partial\mu-\beta^{a}\partial_{a} )
 (1 - \mu\partial
/\partial\mu) \ln Z \nonumber \\
&=&
 (1 - \mu\partial
/\partial\mu)
(\mu\partial /\partial\mu-\beta^{a}\partial_{a} )
 \ln Z \nonumber \\
&=& (1 - \mu\partial /\partial\mu)
(\mu \beta h)
= 0
\,.
\ena
That is, the entropy is not sensitive to a shift of the ground state
energy.  The covariance of $\beta^{a}$ is just its
$\mu$-independence.  The metric $g_{ab}$ is covariant under
renormalization because it is defined in terms of normalized,
connected correlation functions, in Eq.~\ref{metric}.

To show that the metric $g_{ab}$ is positively definite,
we need only remark
that $g_{ab}\delta\lambda^{a}\delta\lambda^{b}$
is given in Eq.~\ref{metric}
as a positive two point function of $\phi=\delta\lambda^{a}\phi_{a}$,
integrated against a positive function.

The cosine term in the metric
plays a twofold role.  On the one hand,
it provides the $\Theta_{bulk}$ term
in the correlation function of $\Theta$
with the boundary operator.
On the other
hand the cosine term
renders the metric independent of contact terms in the
two-point functions of the boundary operators.
Such terms
could spoil the positivity of the metric.
The metric, as defined by Eq.~\ref{metric},
is independent of contact terms.
During the proof of the gradient formula, we split it into two
parts, each of which
does depend on the contact terms.
At that point,
the two point functions have to be treated as distributions.
In the end, when the two terms are joined together,
the result is independent of the contact terms.
The technical roles of the cosine term are evident,
but we do not see a deeper meaning.
The cosine first appeared
in the string theory metric proposed in Ref.~\cite{Witt1}.
But the proposal was not natural in string theory,
as it involved integrating dimension zero fields.
So we still do not see a natural interpretation
of the cosine term.

\section{Relation with string theory}
The conjectured string theory gradient formula
involves
an additional boundary
coupling constant $\lambda^{0}$ which couples to the
identity operator $\phi_{0}=\idop$.
The string partition function is
\[
\Zstr = \exp(\mu\beta \lambda^{0}) z(\mu\beta)
\]
where $z(\mu\beta)$ is the boundary partition
function, from Eq.~\ref{eq:partitionfn}.
The string $\beta$-function, $\betastr^{a}$,
is the ordinary $\beta^{a}$ for the ordinary coupling constants,
plus, from
Eq.~\ref{eq:rgz},
$\betastr^{0}= \lambda^{0} + h(\lambda)$.
The conjectured string theory gradient formula is
\[
\Gstr_{ab} \betastr^{b} = - \partial_{a}\gstr
\]
where
\[
\gstr = (1- \mu\partial/\partial\mu) \Zstr
\]
and the string metric is
\eq
\Gstr_{ab}(\lambda) = \textint\mu \dif \tau_{1}
\textint\mu \dif \tau
\,  \Zstr \langle \phi_{a}(\tau_{1})\phi_{b}(\tau )\rangle
\left (1-\cos\left[2\pi(\tau -\tau_{1})/\beta\right] \right) \, .
\en
These string formulas are un-physical, when applied to 1-d quantum
systems.  No physical probe couples to the identity operator
$\phi_{0}=\idop$, so $\lambda^{0}$ is not a physical coupling
constant.  Un-normalized correlation functions are not measurable.
Changes in $\gstr$ are not locally measurable, because $\gstr$ is
constructed from $z$, not $\ln z$.  On the other hand, all of the
elements of the physical gradient formula, Eq.~\ref{eq:gradient},
can by measured by local operations at the boundary of the 1-d
system.  The string gradient formula
\emph{is} formally sensible from the string theory perspective.  The
$\lambda^{a}$ are the wave-modes of spacetime fields, $\lambda^{0}$
is the zero-mode of the tachyon field.  The equation
$\betastr^{a}=0$ has the form of a space-time equation of motion.
The function $\gstr(\lambda)$ has the form of a space-time action.

The un-physical parameter $\lambda^{0}$ can be eliminated
by extremizing $\gstr$\cite{KMM,Moor1}.
We carry out this idea.  We calculate that
$\partial \gstr/\partial \lambda^{0} =0$ at
$\lambda^{0}=a_{0}$,
$\mu \beta a_{0} = - \mu \partial \ln z/\partial \mu$.
We calculate that,
at $\lambda^{0}=a_{0}$,
$\gstr = \Zstr = z \exp(\mu\beta a_{0})$,
which is the physical quantity $\exp(s)$.
It now becomes
straightforward to show the equivalence between the string
gradient formula and the physical gradient formula.
The string gradient formula is trivial
in the direction of $\lambda^{0}$,
and is precisely the physical formula
on the subspace $\lambda^{0}=a_{0}$.
To be explicit,
the components of the string metric are
$\Gstr_{00}=(\mu\beta)^{2}\Zstr$,
$\Gstr_{0b} = \mu\beta \Zstr \partial_{a}\ln \Zstr$,
$\Gstr_{ab} = \Zstr (g_{ab} +
\partial_{a}\ln \Zstr \partial_{b} \ln \Zstr)$,
where the indices $a,b$ now range only over the physical coupling
constants.
The string gradient formula
splits into two equations
\eqa
\Gstr_{00}\betastr^{0} + \Gstr_{0b}\beta^{b}  = -\partial_{0}\gstr
\nonumber \\
\Gstr_{a0}\betastr^{0} +\Gstr_{ab}\beta^{b}  = -\partial_{a}\gstr
\nonumber
\,.
\ena
The first is satisfied identically,
it is just the rg equation for $\Zstr$,
\[
\mu\partial\Zstr /\partial\mu =
(\beta^{a} \partial_{a}+\mu\beta \betastr^{0})\Zstr
\,.
\]
The second equation, after substituting and
then using the rg equation for $\Zstr$,
becomes
\[
\Zstr g_{ab}\beta^{b} =
\Zstr (-1 + \mu\partial/\partial\mu)
\partial_{a}\ln \Zstr
\]
which is exactly the physical gradient formula,
since $\partial_{a}\ln \Zstr=\partial_{a}\ln Z$.
So, by proving the physical gradient formula,
we have also proved the string gradient formula.

We would like to thank Gregory Moore and Alexander Zamolodchikov for
stimulating discussions.  We especially thank Gregory Moore for
pointing out a deficiency in an early version of the proof.  This
work was supported by the Rutgers New High Energy Theory Center,
which A.~K. thanks for warm hospitality.  The work of A.~K. was
supported in part by BSF-American-Israel Bi-National Science
Foundation, the Israel Academy of Sciences and Humanities-Centers of
Excellence Program, the German-Israel Bi-National Science
Foundation.

\bibliography{boundary_entropy}
\end{document}